\pdfoutput=1

\documentclass[letterpaper,twocolumn,10pt]{article}
\usepackage{usenix-2020-09}

\usepackage{amsmath}
\usepackage{amssymb}
\usepackage{nicefrac}

\usepackage{tikz}
\usepackage{tikz-network}
\usepackage{pgfplots}
\usepackage{bytefield}
\usepackage{subfigure}

\usepackage{makecell}

\usepackage{xurl}

\usepackage[ruled,vlined,linesnumbered]{algorithm2e} 
\SetKwInput{KwInput}{Input}                
\SetKwInput{KwOutput}{Output}    

\usepackage[capitalize,noabbrev,nameinlink]{cleveref}
\usepackage{apptools}

\usepackage{listings}
\usepackage[frozencache=true,cachedir=minted-cache]{minted}

\usepackage{booktabs}
\crefformat{section}{§#2#1#3}
\crefformat{subsection}{§#2#1#3}

\newcommand{\OMIT}[1]{}

\newcommand{\EG}{\emph{e.g.}\xspace}
\newcommand{\IE}{\emph{i.e.}\xspace}

\newcommand{\sysname}{\textrm{{NetCov}}\xspace}

\newcommand{\paraspace}{\vspace{1ex}}
\newcommand{\para}[1]{\paraspace\noindent\textbf{#1.}\xspace}

\usepackage[framemethod=tikz]{mdframed}
\usepackage{enumitem}
\usepackage[ampersand]{easylist}
\usepackage{graphicx}
\usepackage{pbox}
\usepackage{multirow}
\usepackage{etoolbox}

\definecolor{red}{rgb}{0.6,0,0} 
\definecolor{blue}{rgb}{0,0,0.6}
\definecolor{green}{rgb}{0,0.8,0}
\definecolor{cyan}{rgb}{0.0,0.6,0.6}
\definecolor{lightgray}{gray}{0.98}
\definecolor{lightblue}{rgb}{0.13, 0.67, 0.8}
\definecolor{lightorange}{RGB}{255,247,230}
\definecolor{codegreen}{rgb}{0,0.6,0}
\definecolor{codegray}{rgb}{0.5,0.5,0.5}
\definecolor{codepurple}{rgb}{0.58,0,0.82}
\definecolor{keywordcolor}{RGB}{94,20,64}
\definecolor{bluekeywords}{rgb}{0,0,1}
\definecolor{greencomments}{rgb}{0,0.5,0}
\definecolor{redstrings}{rgb}{0.64,0.08,0.08}
\definecolor{xmlcomments}{rgb}{0.5,0.5,0.5}
\definecolor{types}{rgb}{0.17,0.57,0.68}
\definecolor{KWColor}{RGB}{0,0,255}
\definecolor{AnnotationColor}{RGB}{0,137,180}
\definecolor{CommentColor}{rgb}{0.12,0.38,0.18}
\definecolor{StringColor}{rgb}{0.06,0.10,0.98}
\definecolor{darkred}{rgb}{0.65,0,0}
\definecolor{lightgrey}{rgb}{0.8,0.8,0.8}
\definecolor{marmalade}{RGB}{193,101,18}
\definecolor{peach}{RGB}{250,217,193}
\definecolor{lime}{RGB}{220,237,193}

\lstdefinestyle{VisualStudio}{
  xleftmargin=0pt,
  basicstyle=\ttfamily\small,
  commentstyle=\color{CommentColor}\ttfamily\small,
  stringstyle=\color{darkred},
  keywordstyle=\color{KWColor},
  escapeinside={/*@}{@*/}
}

\newcommand{\code}[1]{\lstinline{#1}}

\usepackage{authblk}
\makeatletter
\renewcommand\AB@affilsepx{~ \protect\Affilfont} 
\makeatother
\begin{document}

\date{}

\title{\Large \bf Test Coverage for Network Configurations}



\author[1]{Xieyang Xu}
\author[1]{Weixin Deng}
\author[2]{Ryan Beckett}
\author[1,3]{Ratul Mahajan}
\author[4]{David Walker}
\affil[1]{\textit{University of Washington}}
\affil[2]{\textit{Microsoft}}
\affil[3]{\textit{Intentionet}}
\affil[4]{\textit{Princeton University}}

\maketitle

\textbf{Abstract---}
We develop \sysname, the first tool to reveal which network configuration lines are being tested by a suite of network tests. It helps network engineers improve test suites and thus increase network reliability.  A key challenge in its development is that many network tests test the data plane instead of testing the configurations (control plane) directly. We must be able to efficiently infer which configuration elements contribute to tested data plane elements, even when such contributions are non-local (on remote devices) or non-deterministic.
\sysname uses an information flow graph based model that precisely captures various forms of contributions and a scalable method to lazily infer contributions.
Using it, we show that an existing test suite for Internet2 (a nation-wide backbone network in the USA) covers only 26\% of the configuration lines. The feedback from \sysname makes it easy to define new tests that improve coverage. For Internet2, adding just three such tests covers an additional 17\% of the lines.

\section{Introduction}
\label{sec:introduction}

As critical infrastructure, networks must be highly reliable but, unfortunately, network outages are quite common. A primary culprit is networks' reliance on complex, low-level commands embedded in their configuration. The configurations dictate how routers select best paths and forward traffic.
Day-to-day updates to them is error-prone, leading to outages that knock off important online services (e.g., banking), ground airplanes, and disable critical communication (e.g., emergency calls)~\cite{time-warner,united,twitter,bank, t-mobile}.


To improve network reliability, automatic testing and verification of configurations is becoming commonplace.
Today, network operators have at their disposal many tools with increasing sophistication that can scale to large networks and check various aspects of network behavior \cite{rcdc,jingjing,hoyan,libra,aws-inspector}. 

However, using such tools is not sufficient by itself; one must also use them \textit{effectively}. Outages can occur despite automated testing when the test suite is poor and does not cover key aspects of network configuration. This was the case with the massive Facebook outage during which Facebook, WhatsApp, Instagram, and Oculus were unavailable for six hours~\cite{fboutage}. 
%
Current tools have pushed the limits of \textit{what can be tested} but left open the question of \textit{what need be tested}. 

Without tool support, it is difficult for engineers to know if they are effectively testing network configurations. 
In industrial networks with hundreds of thousands of lines of configurations, engineers' understanding of network behavior and dependencies is necessarily incomplete. It is even harder to evolve an existing test suite after the network evolves because the engineers likely do not know what the old test suite is or is not testing for the updated network. 

Recent work has proposed data plane coverage~\cite{yardstick} to reveal testing gaps. It shows which data plane elements, such as forwarding rules, are exercised by a test suite.
However, well-tested data plane does not imply well-tested configurations.   
Data plane elements are the output of network's configurations (which define its control plane) and the current operating environment (failures, external routing information). 
Testing a given data plane only tests configuration elements that  are exercised in that particular environment.
Other configuration elements are not tested. We demonstrate this empirically via a scenario where testing {\em all} data plane elements leaves over half of configuration lines untested.



We develop \textit{configuration coverage} to provide comprehensive and precise feedback to network engineers on test suite quality. Our goal is to identify exactly which configuration lines are being tested and which ones are not. Further, we want to consider all configuration elements, not only those that contribute to the current data plane. The precise nature of this feedback (untested configuration lines) helps improve tests---add tests that target untested lines---which in turn can improve network reliability. This is similar to how code coverage tools help improve tests and software quality~\cite{google-coverage, vs-coverage, codecov}.

A major challenge we face is that many network tests do not exercise configurations directly. Instead, they reason about the data plane elements produced by configurations. We need to infer the configuration elements that contribute to the a tested data plane element. 
This inference is complicated because contributions can be non-local and non-deterministic. In a distributed control plane, a piece of tested routing information may have been propagated and transformed multiple times along its path, and both local and non-local configurations may have contributed to its existence. For example, the path attributes of a BGP route is shaped by routing policies on each and every hop that it traversed. Further, not all contributions are deterministic. For instance, any one of possibly multiple sub-prefixes can lead to the route of an aggregate prefix. We must scalably account for local and non-local contributions and for non-deterministic contributions.

Our solution is to model the contribution between configuration elements and data plane elements as an \textit{information flow graph}. An IFG is a directed acyclic graph (DAG) where vertices denote network elements and edges denote contributions. In addition to direct contributions from configuration elements to data plane elements, we also model contributions between data plane elements (from predecessors to successors). For instance, a BGP route contributes to the BGP message that derived from it. Indirect contributions are thus modeled by multi-hop paths in the DAG. 
When contributions exhibit non-determinism, we use special \textit{disjunctive} nodes to organize possible DAG paths that may contribute to a given data plane element.

We build a tool called \sysname based on this model. It annotates which configuration lines and logical elements are tested by a given test suite and produces aggregated coverage statistics.  
To efficiently map tested data plane element to the set of contributing configuration elements, it materializes the IFG lazily, instead of tracking contributions proactively (during data plane generation). This design avoids the cost to compute and store contributions for transient or untested data plane elements. 
\sysname is open-sourced on GitHub~\cite{netcov-github}.



We evaluate \sysname on Internet2, a nation-wide backbone network in the USA, and on synthetic data center networks. We show that test suites proposed in prior work can have poor coverage. For instance, we found that the three tests proposed by Bagpipe~\cite{bagpipe} covered only 26\% of the configuration lines of Internet2. 
We also show how surfacing untested configuration elements suggests new tests that will improve coverage. By adding just three such tests to the Internet2 test suite based on \sysname's feedback, we could improve coverage to 43\%, and more similar tests can be added to further increase coverage.
%
\sysname performs reasonably well. The time to compute coverage is 1.2 hours for the largest network that we study, which has over 2 million forwarding rules. This time is an order of magnitude less than the time to execute tests. 

Stepping back, we note that networking is not alone in its reliance on configuration. Today, a lot of infrastructure and distributed applications are deployed by composing existing components using configuration (e.g., infrastructure deployment using Terraform, and application deployment using containers and service meshes). These  configurations are central to correct behavior, which is why there is an intense focus on testing them properly~\cite{terraform-test, istio-test}. As for networks, there are no tools to help engineers discover how well the configurations are tested. The techniques developed in our work, the IFG-based contribution tracking and its lazy traversal, can provide a starting point toward better testing of infrastructure and distributed application configuration as well.

\section{Background on Network Testing} 
\label{sec:background}

In networks with distributed control planes, each device runs one or more routing protocol (e.g., BGP, OSPF) instances. Each instance exchanges routing messages with its neighboring instances. Routing messages contain attributes of paths that the sender is using to various destinations. A routing instance may learn multiple paths to the same destination via different neighbors. It selects the best one (or multiple best ones if multipath routing is enabled) based on its policy and stores that path in its protocol RIB (routing information base). 
Multiple routing protocol instances on a device may have best paths to the same destination. The device selects the best one(s) based on the relative preference of the protocols and stores the selection in its main RIB. 
Information in the main RIB is used to forward packets.\footnote{In reality, for fast forwarding, routers have a forwarding information base (FIB), which maps each main RIB destination to its outgoing interface, by recursively resolving next hop information (which may be an IP address). The difference between main RIB and FIB is not material for our work, and we use the term main RIB for the table that has forwarding information.}

Network engineers can control many aspects of the computation above using device configuration. This includes the routing protocol instances that are running; the peering between instances; the destination prefixes that are announced by each routing protocol instance; how routing messages are transformed prior to sending (export policy) and upon reception (import policy); and the preference function for best path selection. Naturally, thus, how the network forwards packets is intimately dependent on device configurations.

Given the importance of configurations to correct network behavior, network engineers use automatic testing and verification to find bugs and gain confidence in their correctness. Network tests come in two flavors. {\em Data plane tests} analyze the computed data plane state (\IE, RIBs), e.g.,
checking that node A can reach B and that route to a particular destination is present at node C. {\em Control plane tests} directly analyze device configuration, e.g., checking  that the import policy blocks routing messages for private address space (such as 10.0.0.0/8) and BGP peerings are correctly configured.

\section{Configuration Coverage: Overview}\label{sec:overview}

Network engineers today create data and control plane tests based on past outages and their knowledge of which behaviors are important to test. There are no tools to provide feedback on how well they are testing configurations and which aspects of the configuration are untested. We aim to build such a tool. Given the complexity of real-world networks, it is difficult for humans to know if they have covered all important elements of configurations.
As with software, high coverage is necessary but not sufficient for a good test suite.  In addition to exercising all key behaviors, the tests must also properly assert that those behaviors match intent. This latter task is not our focus.

We now outline how we compute which configuration elements are covered by a suite of data and control plane tests. 

\subsection{Defining coverage} 

We deem a configuration element to be covered if it $i)$ is tested directly by a control plane test; or $ii)$ contributes to the production of a data plane element tested by a data plane test. 
For now, assume that contributions are deterministic. We discuss non-deterministic contributions in the next section.

\begin{figure}[t!]
  \centering
    \includegraphics[width=\columnwidth]{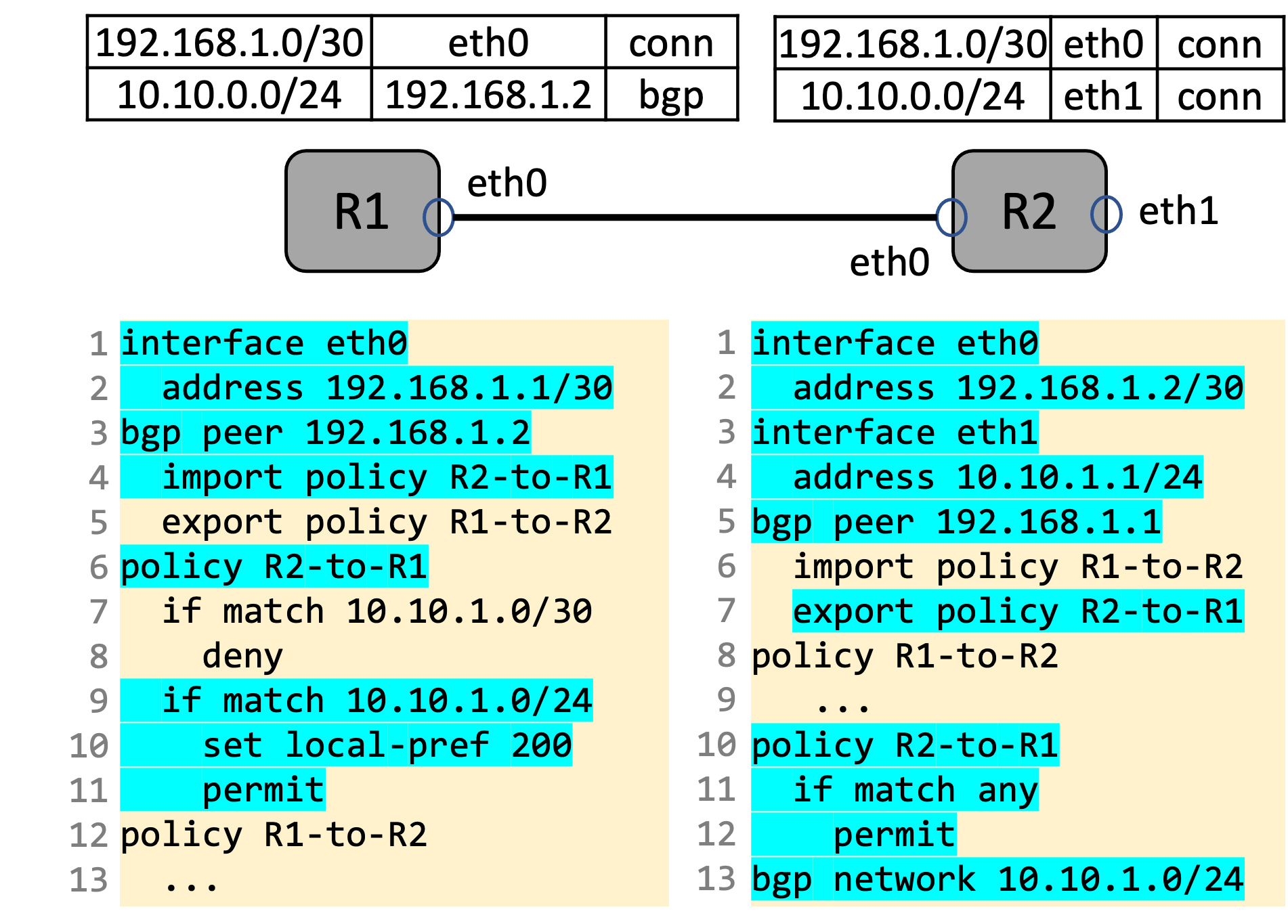}
    \caption{An example network with routing tables and configurations. The highlighted configuration lines are covered when the route to 10.10.1.0/24 is tested at R1.}
    \label{fig:coverage-example}
\end{figure}

\autoref{fig:coverage-example} illustrates configuration coverage as a result of a data plane test. It shows parts of the two routers' configuration. R1's configuration defines one interface (Lines 1-2) and one BGP peer (192.168.1.2, which is R2's address), and it specifies the import and export policy to use. The import policy (R2-to-R1 at Lines 6-11) denies routing messages for a particular prefix and sets the preference for another. 

R2's configuration defines two interfaces, a BGP peer (R1) and routing policies. At Line 13, it states that the prefix 10.10.1.0/24 should be announced to BGP peers {\em iff} it is in the main RIB.\footnote{Different router vendors have different semantics for BGP network statements. We are assuming Cisco semantics.} In our example, 10.10.1.0/24 will be in the main RIB as it corresponds to the eth1's prefix. (Address statements like Line 4 encode the IP address and prefix length. For eth1, given the address 10.10.1.1 and prefix length of 24, the prefix is 10.10.1.0/24.)
Routers add interface prefixes to the "connected" protocol RIB, from where those prefixes can enter the main RIB. The resulting RIBs on the two routers are shown in the figure. Each entry includes the next hop and source routing protocol ("conn" = connected). 

Suppose the entry for 10.10.1.0/24 at R1 was tested by a data plane test. The covered configuration elements are highlighted. On R1, the BGP peer configuration and import policy binding (Lines 3-4) are covered because the tested entry came via that peering and passed through that policy. Parts of the routing policy R2-to-R1 relevant to the tested state (Lines 6, 9-11) are also covered. The interface definition (Lines 1-2) is covered because it enables the BGP peering to be established. In contrast, the export policy R1-to-R2 and unexercised parts of R2-to-R1 (Lines 7-8) are not covered. 

There are covered configuration elements at R2 as well. These include the interface definitions---eth0 enables the BGP edge and 10.10.1.0/24 was announced due to eth1---and BGP peering, the export policy, and the BGP network statement.

\para{Alternative definitions of coverage}
One may consider an alternative definitions of coverage that disregards non-local configuration elements. But we posit that including non-local elements is more meaningful. These elements, such as the BGP network statement on R2's Line 13, are just as key to the existence of 10.10.1.0/24 at R1 as the local elements. 

Another definition of coverage is based on mutation~\cite{mutation-coverage}: a configuration line is covered if its mutation alters the test result. 
This definition will report an additional class of configuration elements as covered---configuration elements that de-prioritize (or reject) the competitors of the tested data plane element. 
Mutation-based coverage tends to be significantly harder to compute~\cite{mutation-survey}, and its results can be hard to interpret. In developing the first tool in this space, we decided to focus on a simpler, more direct definition of coverage. We will explore more sophisticated definitions in the future.




\subsection{Our approach}
While it is straightforward to identify configuration elements covered by a control plane test, it is not for data plane tests. 
Data plane tests analyze the "output" of the control plane, and we need a scalable way to compute which configuration elements contributed to tested data plane state. The relationship between these inputs and outputs is complex. How a particular RIB entry comes about relies on many configuration elements across multiple devices. 
The need to map tested outputs to input space sets computation of configuration coverage apart from data plane coverage and software coverage, for both of which the coverage domain is the same as test domain.



To motivate our approach to solving this problem, let us first sketch two strawman approaches. 
One potential approach is to express control plane computation  declaratively, e.g., in Datalog. This enables identification of contributing inputs for a given output using a form of backward-reasoning~\cite{positiveprovenance,negativeprovenance}.
However, network control plane computations can be quite complex (e.g., non-monotonic behaviors \cite{stablepath, pathvector}). While declarative encodings may work in special cases~\cite{fastplane}, it is hard to get high-fidelty, performant encodings for the general case. 
That is why most control plane analysis tools use an imperative approach~\cite{hoyan,batfish,plankton,cbgp}.\footnote{Batfish~\cite{batfish}, a widely used control plane analysis tool, originally used Datalog to encode network control planes but switched to imperative simulations due to expressiveness and performance challenges.}

Another potential approach is to use simulation-based forward reasoning, i.e., simulate the control plane (imperatively) and track which configuration elements feed into each part of the data plane state. 
However, this approach has scalability limitations.  Network simulation is time and memory intensive~\cite{batfish,hoyan,cbgp}, and it will become significantly worse if it needed to track all necessary information along each hop. 


\begin{figure*}[t!]
  \centering
    \includegraphics[width=0.9\linewidth]{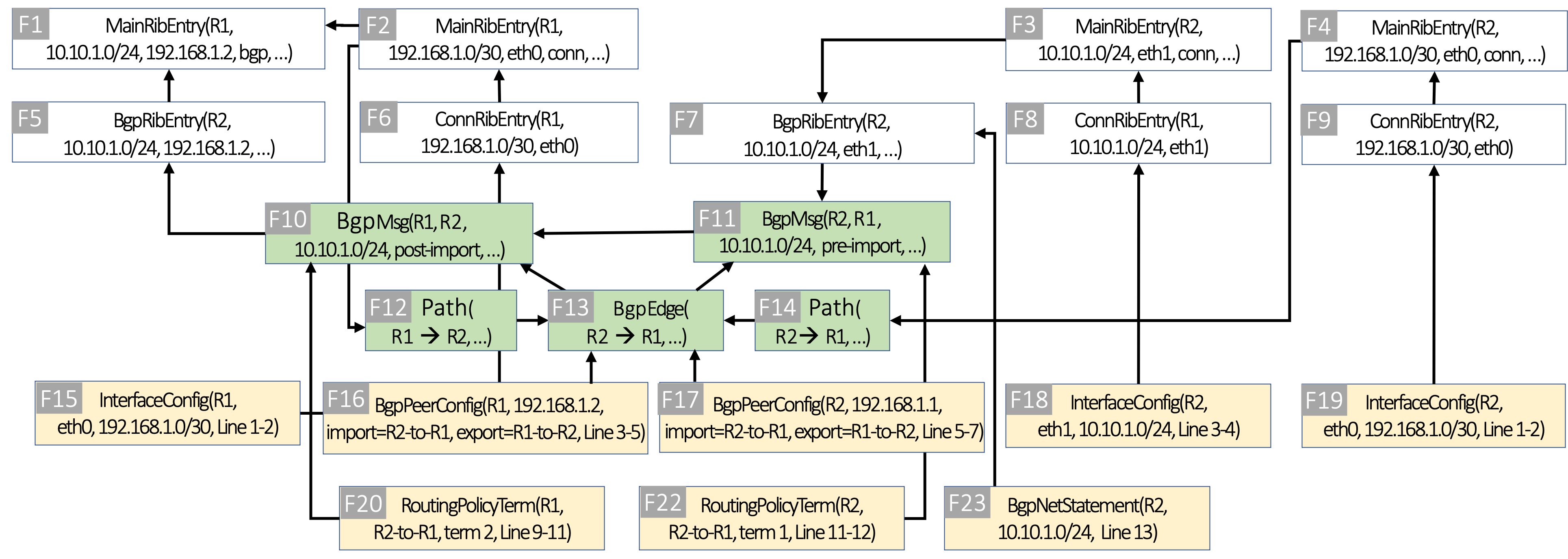}
    \caption{Subset of IFG for the \autoref{fig:coverage-example} example, to track configuration elements  on both routers that contribute to the tested RIB entry (F1). Colors denote different fact types: data plane state (white), configuration elements (yellow), auxiliary facts (green).}
    \label{fig:coverage-graph}
\end{figure*}

Our approach is based on two observations. First, for the purposes of computing coverage, we do not need a full computational model of the control plane. We need to only track which configuration elements contribute to tested data plane state (i.e., taint analysis~\cite{taj}), not the exact input-output relationship; and we need to reason only about the stable state (i.e., the the of devices once they have settled on best paths), not the transient states. Data plane testing~\cite{hsa,veriflow,deltanet,atomic,rcdc} assumes that the analyzed state is stable.
%
Our second observation is that the stable state contains enough information for us to infer contributions of configuration elements after the fact, based on the semantics of the control plane. This inference is vastly cheaper than tracking contributions towards all data plane state entries, independent of whether they are tested.



To model and infer contributions to the stable state, we use an {\em information flow graph} (IFG). 
\autoref{fig:coverage-graph} shows a simplified subset of IFG for the \autoref{fig:coverage-example} example. Each node is a {\em fact} and each arrow denotes direct information flow from the tail to head. IFGs have three types of facts: $i)$ data plane state, $ii)$ configuration elements; and $iii)$ auxiliary facts that help connect the previous two types. 

The main RIB entry 10.10.1.0/24 at R1 (F1) is derived from the corresponding BGP RIB entry (F5),
which in turn is derived from the BGP message from R2 (F10). This message exists because of the BGP edge between R1 and R2 (F13), the source message sent by R2 (F11), and the relevant configuration element within import policy (F20). R2 sent the BGP message because of the same BGP edge (F13), its export policy elements (F22), and the BGP RIB entry (F7). This BGP RIB entry exists because of the configuration element (F23) and the RIB entry (F3), which exists because of the connected route (F8). The BGP edge (F13) exists because of the configuration elements that define the peering (F16, F17) and paths between R2 and R1 that enable the BGP session to be established. The paths depend on the RIB entries (F2 and F4, respectively), the contributions to which can be similarly traced. In this manner, the IFG captures all configuration elements that led to the tested RIB entry (F1).

We do not track IFG dependencies proactively but infer them on-demand based on control plane semantics, using a mix of backward-forward reasoning. 
Backward inference infers the parent (tail) of the edge from its child (head).
The information in child nodes is not enough to fully recover the parent nodes, but is often enough to select them from the known stable state. For instance, we can compute the BGP RIB entry F5 from the main RIB entry F1---the main RIB entry indicates that its source routing protocol is BGP, and we thus look up the BGP RIB for 10.10.1.0/24.



Lookup-based inference does not always work. For instance, given a BGP message which has passed through an import policy, we can not compute backwards which policy terms of the import policy were exercised (F10 $\leftarrow$ F20). Another parent of F10, the pre-import BGP message (F11) cannot be looked up either because it is not part of the input and needs to be computed on-the-fly. 
To address these limitations, we combine backward and forward inference. When a parent can not be directly looked up, we first look up the prerequisites of the parent. For instance, we can look up F7 based on F10. Next, we use targeted simulations to compute non-existing facts and to select relevant facts exercised in a control plane process or data plane process. For instance, given the BGP route at R2 (F7), we simulate its processing through the export policy, which allows us to derive the pre-import BGP message (F11) and find the policy term exercised during the export process (F22). Once F11 is computed, we conduct another targeted simulation to discover the policy term exercised in the import process (F20).
Unlike a full control plane simulation, these targeted simulations are fast. They have limited scope (e.g., best path selection is not simulated) and are done only for messages of interest, not all messages.

By combining backward and forward inference, atop the stable state IFG, we can scalably discover all covered configuration elements. We describe this approach in detail next.

\section{Design of \sysname}

\sysname takes as input configuration files, data plane state (protocol RIBs, main RIB and active routing edges) of the network. The data plane state may be pulled from live network or produced by a control plane analysis tools~\cite{cbgp,batfish,hoyan}. In addition, \sysname takes as input what is tested: data plane entries that are tested by data plane tests, and configuration elements that are tested by control plane tests.
This information is produced by network testing tools~\cite{yardstick,batfish}. 

Based on these inputs, \sysname computes which configuration elements are covered.
The core of this computation efficiently mapping a data plane fact to configuration elements that contribute to it. We describe this computation next. 


\subsection{Information flow model}\label{sec:ifg}


IFGs are directed acyclic graphs whose nodes denote network \textit{facts} and edges denote information flow between facts. 
\autoref{fig:model} shows the types of network facts modeled by \sysname and the information flow between different types.

Our model has thee types of facts: configuration elements, data plane state, and auxiliary facts. Data plane state has three subtypes: main RIB entries, protocol RIB entries, and access control list (ACL) entries. 
%
%
Auxiliary facts help concisely capture information flow dependencies from configurations to data plane state. They have three subtypes: routing edges, routing messages, and paths that carry routing messages. Routing message facts represent messages between routing protocol instances across devices as well as within a device, i.e., redistribution~\cite{redistribution}. This uniform treatment is a modeling convenience. In reality, explicit messages are not exchanged during redistribution (though redistribution is subject to routing policies akin to messages between cross-device routing instances). 

The last column of \autoref{fig:model} shows how information flows among different types of facts. A main RIB entry stems from a protocol RIB entry and optionally another main RIB entry (when its next hop is an IP address whose corresponding output interface needs further resolution). A protocol RIB fact stems from a routing message (for protocols such as BGP), a configuration element (for connected interfaces and static routes), a main RIB entry accompanied with a configuration element (such as when a BGP network statement populates a main RIB entry into BGP RIB) or a set of RIB entries  accompanied with a configuration element (for aggregate routes). ACLs facts stem from configuration facts and have no other dependencies. Routing messages stem from a RIB fact or another message (e.g., post-import-policy message depends on pre-import-policy message), and they also depend on routing edges and routing policy configurations. Inter-device routing edges stem from paths that enable sessions to be established and configuration facts that define peerings; Intra-device routing edges stem from configuration facts that define redistribution. Finally, path facts depends on main RIB facts and ACL facts that impact routing traffic along the way. 

\begin{table}[t!]
\begin{tabular}{lll}
\toprule
\multicolumn{2}{l}{\textbf{Network fact} } & \textbf{Information flow} 
\\ \midrule
\multicolumn{2}{l}{Configuration element ($c$)}  & None 
\\ \midrule
\multirow{7}{19pt}{Data plane state}
& Main RIB entry ($f$) & \makecell[cl]{$f_i \leftarrow r_j$ \\ $f_i \leftarrow r_j, f_k$}
\\ \cmidrule{2-3}
& Protocol RIB entry ($r$) & \makecell[cl]{$r_i \leftarrow m_j$ \\ $r_i \leftarrow c_j$ \\ $r_i \leftarrow f_j,c_k $ \\ $r_i \leftarrow \{r_{j_1}, ...\}, c_k$}
\\ \cmidrule{2-3}
& ACL entry ($a$) & $a_i \leftarrow \{c_{i_1}, ... \}$ 
\\  \midrule
\multirow{5}{19pt}{Aux-iliary}
& Routing message ($m$) & \makecell[cl]{$m_i \leftarrow r_j, e_k, \{c_{l_1}, ... \}$ \\ $m_i \leftarrow m_j, e_k, \{c_{l_1}, ... \}$ }
\\ \cmidrule{2-3}
& Routing edge ($e$) & \makecell[cl]{$e_i \leftarrow \{c_{j_1},... \}$  \\ $e_i \leftarrow \{c_{j_1},... \}, \{p _{k_1},... \}$ }
\\ \cmidrule{2-3}
& Path ($p$) & $p_i \leftarrow \{f_{j_1}, ...\}, \{a_{k_1}, ...\}$ 
\\ \bottomrule
\end{tabular}
\caption{Information flow model: Types of facts and all possible dependencies for each type. $\{t,...\}$ denotes a set of facts. }
\label{fig:model}
\end{table}

\subsection{Inferring the IFG on demand}\label{sec:lazyconstruction}


Based on the information flow model, \sysname uses a backward-forward inference framework to lazily materialize the IFG from any set of facts whose coverage need to be tracked. The framework can be abstracted with a set of \textit{inference rules} and an iterative construction algorithm. Each inference rule is function that takes a materialized IFG node as input and materializes a set of its ancestor nodes as well as the edges the allows the ancestors to reach the input node. These nodes and edges will be merged into the materialized IFG by the construction algorithm. 
The implementation of these functions uses one or both of the \textit{lookup-based inference} and \textit{simulation-based inference}. Let us elaborate.

\para{Lookup-based inference} 
The computation of a control plane is lossy. For instance, while a main RIB entry may be derived from a BGP RIB entry, we cannot infer the complete BGP RIB entry from the main RIB entry because BGP specific attributes (e.g., AS-path) are not preserved in the main RIB.

\begin{figure}
    \centering
 \begin{minted}[breaklines,linenos,fontsize=\small]{python}
def infer_from_main_rib_entry(f, stable_state):
  if not (f is MainRibEntry and f.protocol == 'bgp'):
    return []
  bgp_entry = stable_state.bgp_rib.lookup(
    host=f.host,
    prefix=f.prefix,
    nexthop=f.nexthop,
    status='BEST'
  )
  return [(bgp_entry, f)] 
 \end{minted}
 \vspace{-15pt}
    \caption{Rule to infer BGP RIB entry from main RIB entry.\label{alg:main-from-bgp}}
\end{figure}

To address this challenge, our inference leverages the stable state. 
It first infers a subset of attributes based on control plane semantics. This partial inference provides enough information for us to look up the complete entry in the stable state.

\Cref{alg:main-from-bgp} shows the simplified function to infer the BGP RIB entry that led to a main RIB entry. 
Based on control plane semantics, if a main RIB entry indicates its source protocol to be BGP, it must have stem from a BGP RIB entry on the same router with the same \code{prefix} and \code{nexthop} attributes (Lines 5-7). Besides, the BGP RIB entry should have been selected as the best route (Line 8). Such information is enough to uniquely identify the parent within the known stable state. The return value (Line 10) is a list of tuples denoting the IFG edges materialized by this rule.


\para{Simulation-based inference} Lookup-based inference is not enough to materialize the IFG. Some facts are not present in the stable state (e.g., routing messages), and some facts do not contain enough information to uniquely identify their parents. We use local simulations to complement lookup-based inference. But simulations can only be performed in the forward direction, \IE, to compute a fact using simulations, we first need to know its parent. 
We use a generalized version of lookup-based inference to discover grandparent facts of a known fact, and then use simulations with the grandparents to infer their children (i.e., parents of the original fact).

\begin{figure}
 \begin{minted}[breaklines,linenos,fontsize=\small]{python}
def infer_from_bgp_message(m, stable_state):
  if not (m is BgpMsg and m.is_post_import):
    return []
  bgp_edge = stable_state.bgp_edges.lookup(
    recv_host=m.host
    send_ip=m.nexthop
  )
  origin_entry = stable_state.bgp_rib.lookup(
    host=bgp_edge.send_host,
    prefix=r.prefix,
    status='BEST'
  )
  pre_import_msg, export_clauses = policy_simulation(
    input=origin_entry,
    policy=bgp_edge.export_policy
  )
  _, import_clauses = policy_simulation(
    input=pre_import_msg,
    policy=bgp_edge.import_policy
  )
  return [(pre_import_msg, m), (bgp_edge, m)] +
    [(cl, m) for cl in import_clauses] + 
    [(origin_entry, pre_import_msg), (bgp_edge, pre_import_msg)] +
    [(cl, pre_import_msg) for cl in export_clauses]
 \end{minted}
\caption{Rule to infer ancestors of a post-import BGP message.}\label{alg:bgp-message}
\end{figure}

\Cref{alg:bgp-message} shows the simplified inference rule that infers the ancestors of a post-import BGP message. Line 13 demonstrates the use of simulation-based forward inference to compute a missing parent fact on the fly. The two prerequisites to simulate the BGP message--the grandparent BGP RIB entry (\code{origin_entry}) and the BGP edge--are discovered via lookup-based backward inference, on Line 8 and Line 4 respectively. The simulation returns the derived BGP message after applying the routing policy, as well as the policy clauses exercised during the process. The second forward-simulation (Line 17) is to discover the policy clauses that are hit during the import process. The return value includes the inferred IFG edges that connect to the input node \code{m} as well as ones that connect to parent \code{pre_import_msg}. The former corresponds to information flow $m_i \leftarrow m_j, e_k, \{c_{l_1},...\}$ in \Cref{fig:model} and the latter corresponds to $m_i \leftarrow r_j, e_k, \{c_{l_1},...\}$.

\para{IFG construction} Next, we detail IFG materialization using inference rules.
Assume for now that the information flow is deterministic; the next section discusses how we handle non-determinism. 

As shown in \Cref{alg:ifg-materialization}, the IFG initially contains only the nodes representing the tested data plane state facts from the input and does not have any edges (Line 2). It is then iteratively expanded by applying inference rules on existing nodes. In each iteration, all inference rules are applied to the dirty nodes derived from the previous iteration (Line 8). The new nodes and edges inferred during such process are collected and merged (with deduplication) into the IFG (Line 9-14). The computation repeats until no new facts can be derived in an iteration.
\begin{algorithm}[t]
\footnotesize
\DontPrintSemicolon
  \SetKwFunction{BuildIFG}{BuildIFG}
  \SetKwProg{myproc}{Procedure}{}{}
  \KwInput{Initial nodes $\{v_i\}$; Inference rules $\{\phi_i: v \mapsto \{(u_i, v_i)\}\}$;}
  \KwOutput{Materialized IFG $(V, E)$}
  \KwData{Stable state data plane state (main RIB and protocol RIBs);
    Routing edges; Configuration elements;}
  
    \myproc{\BuildIFG{$\{v_i\}$, $\{\phi_i\}$}}
    {
        $V, E \leftarrow \{v_i\}, \varnothing$ \;
        $V' \leftarrow \{v_i\}$ \tcp*{dirty nodes of previous iteration}
        \While{$|V'| > 0$}
        {
            $V'' \leftarrow \varnothing$ \tcp*{dirty nodes of current iteration}
            \ForEach{$c \in V'$}
            {
                \ForEach{$\phi \in \{\phi_i\}$}
                {
                    $E' \leftarrow \phi(c)$ \;
                    \ForEach{$(u_i, v_i) \in E'$}
                    {
                        \If{$u_i \notin V$}
                        {
                            $V \leftarrow V \cup \{u_i\}$,
                            $V'' \leftarrow V'' \cup \{u_i\}$
                        }
                        \If{$v_i \notin V$}
                        {
                            $V \leftarrow V \cup \{v_i\}$,
                            $V'' \leftarrow V'' \cup \{v_i\}$
                        }
                        \lIf{$(u_i, v_i) \notin E$}
                        {
                            $E \leftarrow E \cup \{(u_i, v_i)\}$
                        }
                    }
                }
            }
            $V' \leftarrow V''$\;
        }
        \Return{$(V, E)$}
    }
\caption{IFG lazy materialization}\label{alg:ifg-materialization}
\end{algorithm}

\subsection{Handling uncertainty}\label{sec:weak_coverage} There are situations where it is not certain which stable state facts contributes to a given fact. One such scenario is BGP aggregation, where a prefix (e.g., 10.10.0.0/16) is added to the RIB iff at least one more of its more specific prefixes (e.g., 10.10.1.0/24) is present. When multiple more specifics are present, we do not know which one triggered the aggregate. Another such scenario is when multiple paths are available for a routing edge to be established, which can happen when the network uses multipath routing. Here, we do not know which path is actually used by routing messages.

\begin{figure}
  \centering
  \subfigure[\label{fig:uncertainty-model}]{
    \includegraphics[height=.8in]{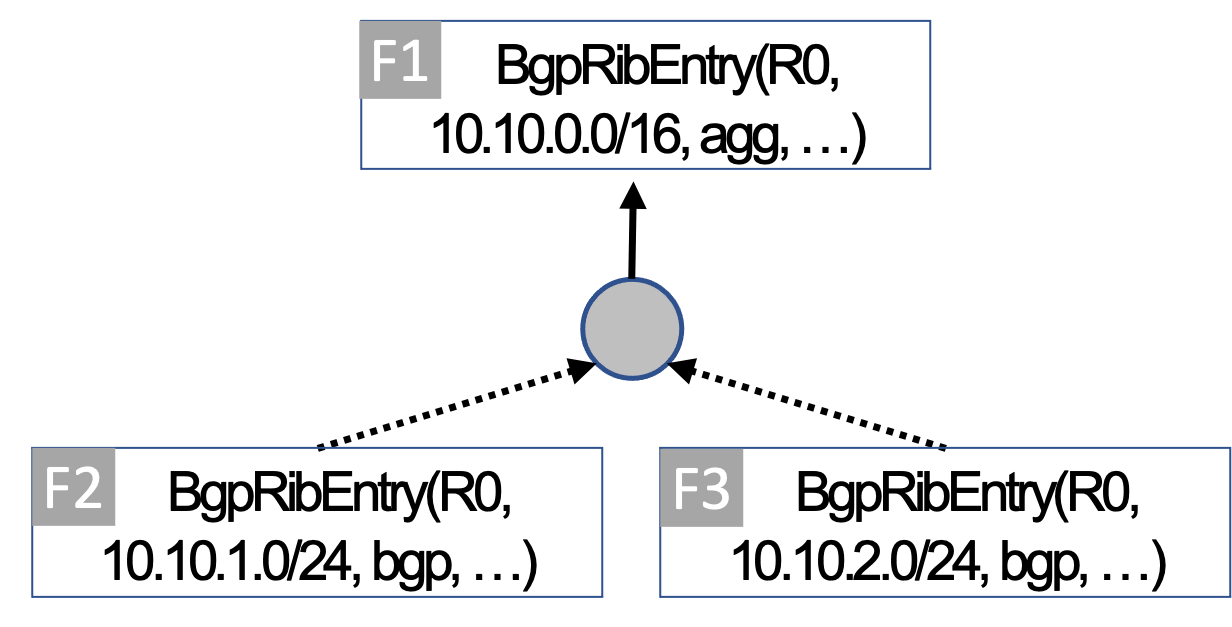}
    }
   \subfigure[\label{fig:weak-coverage}]{
    \includegraphics[height=.9in]{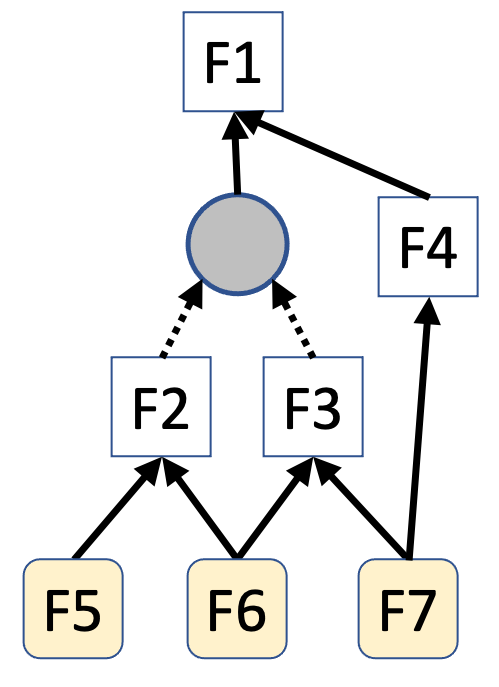}
   }
   \subfigure[\label{fig:bdd-example}]{
    \includegraphics[height=.9in]{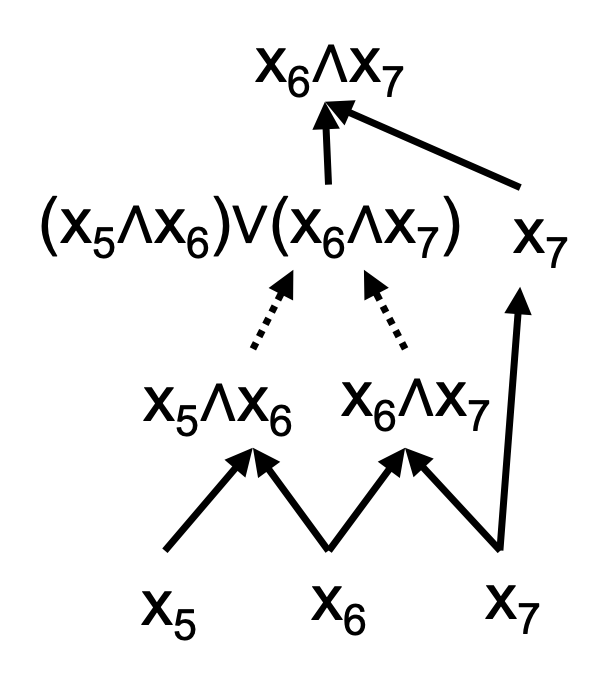}
   }
   \vspace{-10pt}
    \caption{Modeling uncertainty. (a) BGP aggregate (F1) has two potential contributors. (b) F5 is weakly covered but F6 and F7 are strongly covered. (c) The predicates of IFG nodes.}
    \label{fig:uncertainty}
\end{figure}

We model such uncertainty using {\em disjunctive} nodes in the IFG. This node points to the parent fact (e.g., the aggregated RIB fact) and the multiple contributors to the parent point to this node. See \Cref{fig:uncertainty-model} for an example where a BGP aggregate could be triggered by either of the two more specific prefixes.  When our inference rules encounter uncertainty during IFG materialization, they produce a disjunctive node and attach all contributors to it as children.

When computing coverage, we must account for uncertainty because the notion of coverage is different. Assume in \Cref{fig:uncertainty-model} that F1 was tested. If the uncertainty were not there, and F2 and F3 directly pointed to F1, the configuration elements that led to F2 and F3 would have been critical to the outcome. But with the uncertainty, those configuration elements are not critical. The configuration elements that led to F2 could disappear without impacting F1.

To model such possibilities, we introduce the notion of {\em weak} coverage. A configuration element is weakly covered if it contributes to a tested fact but its contribution is not critical. %
In Figure~\autoref{fig:weak-coverage}, assume that F1 is the tested fact. Here, F5 is weakly covered; F1 can be derived without any contribution from F5 because F3 can be derived via F6, which is enough to derive the disjunctive child of F1. 
F6 is strongly covered; without it, neither F2 nor F3 can be derived and thus the disjunctive node cannot be derived. 
F7 is also strongly covered because it contributes to F4, which is essential to F1. 

%

After materializing the IFG, \sysname labels each covered configuration element as strong or weak. The label is determined  
as follows. We first assign a Boolean variable to each configuration element in the IFG. Next, we build a Boolean predicate of each IFG node on top of these variables. The predicate of a fact depends on the predicate of its ancestors in the IFG: A normal node depends on the conjunction of its immediate parents, and a disjunctive node depends on the disjunction of parents. Therefore the predicate of any IFG node is ultimately composed of the variables associated with configuration elements that lead to it, denoted as $\Gamma(v)=F(x_1, \dots, x_n)$. \Cref{fig:bdd-example}
shows the predicates of IFG nodes in \Cref{fig:weak-coverage}. We represent these Boolean predicates using Binary Decision Diagrams (BDDs)~\cite{bdd} and build BDD predicates by traversing the IFG. Once the predicates are built, we test graph reachability and logical necessity between each pair of configuration facts and tested facts. Necessity $\neg x_i \Rightarrow \neg \Gamma(v)$ is equivalent to unsatisfiability of $\neg x_i \wedge \Gamma(v)$. While (un)satisfiability is NP-Complete in general cases, we note that it is efficient in our case---it can be reduced to computing the cofactor $\Gamma(v)|_{x_i=0}$ and testing whether the cofactor is constant false, both of which are efficient using BDD operations.

We further reduce the size of BDD predicates by precluding configuration facts that can reach tested facts via a path with no disjunctive node, such as node F7 in \Cref{fig:weak-coverage}. These configuration facts must be strongly covered so their necessity do not need to be tested. Besides, their validity variables can be replaced with constant true when building BDD predicates, which will not affect the strong/weak classification of other configuration elements. 
We empirically find this heuristic to be effective in reducing the number of variables used for weak coverage computation.







\section{Implementation} \label{sec:implementation}


We implemented \sysname with 4,000 lines of Python code. A total of 18 lambdas (Python functions) encode the IFG inference rules. 
\sysname uses Batfish~\cite{batfish-gh} to extract configuration elements from configuration files and to run targeted simulations, and it uses CUDD \cite{cudd} for BDD operations.


\begin{table}\footnotesize
\centering
\setlength{\tabcolsep}{3pt}
\begin{tabular}{ll}
\toprule
\textbf{Type} & \textbf{Purpose} \\ \midrule
Interface & Interface and its settings (e.g., addresses) 
\\
BGP peer & BGP peer settings (e.g., IP address, AS number)
\\
BGP peer group & BGP peer settings inherited by one or more peers
\\
Route policy clause & One clause in an export or import route policy
\\
Prefix list & List of prefixes, used in route policy clauses
\\
Community list & List of BGP communities for route policy clauses  
\\
AS-path list & List of AS-path expressions for route policy clauses
\\ \bottomrule
\end{tabular}
    \caption{Configuration elements analyzed by \sysname.}
    \label{tbl:config-elements}
\end{table}

\sysname supports all major router vendors supported by Batfish, including Arista, Cisco, and Juniper. It builds a vendor-neutral representation of configuration elements using vendor-specific information provided by Batfish. 
Types of configuration elements currently analyzed by \sysname are listed in \autoref{tbl:config-elements}.  
%


\sysname may not consider all components of a device's configuration. One category of such components device management configuration (e.g., login settings), which does not impact data or control plane functionality. The second category is control plane components that are not currently modeled by \sysname. This includes IPv6 (which is not modeled by Batfish currently) and routing protocols other than BGP (e.g., OSPF). The presence of unconsidered components does not imply that \sysname cannot be used for that network. As we show in the next section, \sysname provides helpful coverage information for parts that are considered.

After constructing the IFG, which yields information on which configuration elements are covered, \sysname computes which lines are covered. Each element typically spans multiple configuration lines, and when an element is covered, it deems all of those lines as covered. 

\begin{figure}[t]
    \subfigure[Line-level coverage. Green background denotes covered lines, and red denotes uncovered lines. Some lines are collapsed for simplicity.\label{fig:screenshot_lcov}]{
        \includegraphics[width=\columnwidth]{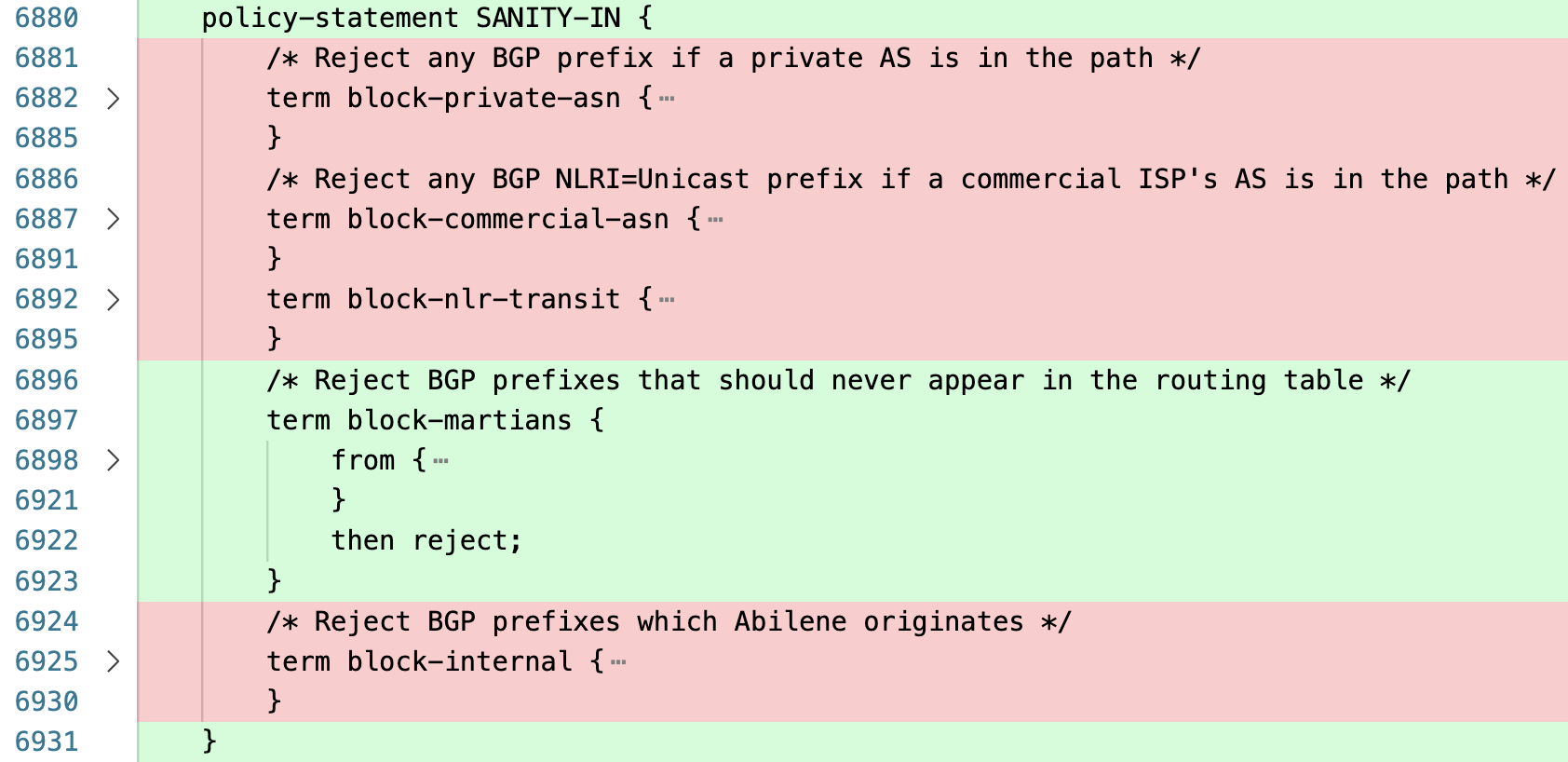}}
    \hfill
    \subfigure[File-level aggregate coverage. The overall coverage is at top right, and the coverage for individual files (devices) is in the table.\label{fig:screenshot_aggregate}]{
        \includegraphics[width=\columnwidth]{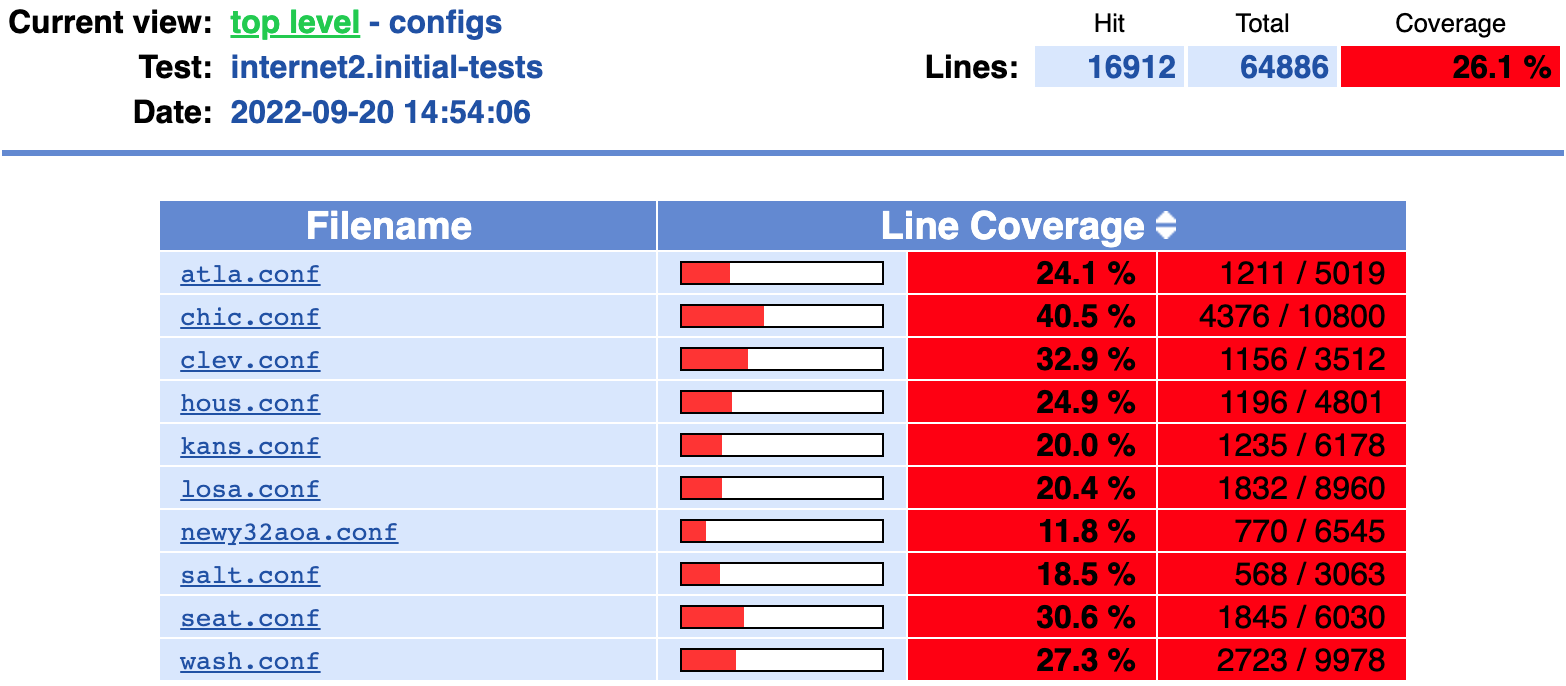}}
    \vspace{-15pt}
    \caption{Example \sysname outputs.} \label{fig:screenshot}
\end{figure}

Based on element and line coverage, \sysname produces three main outputs. The first is a coverage report at the granularity of individual lines (or elements). We produce this report in the \code{lcov} format, which is supported by common code coverage tools and enables users to visualize coverage results as annotations on configuration files. See  \Cref{fig:screenshot_lcov} for an example. The second is coverage aggregated at the file level, generated with the help of GNU LCOV~\cite{lcov}. See \Cref{fig:screenshot_aggregate} for an example. The third output is coverage aggregated by the type of configuration element, which shows what fraction of elements of each type are covered.


These outputs help users uncover testing gaps and improve their test suites in different ways. The aggregate results help identify systematic gaps such as "router A is poorly covered" or "routing policy clauses are poorly covered." The line-level results help them zoom in to specific gaps and develop tests that target them. The case study in the next section demonstrates this test suite improvement process.

\section{Case Studies} 
\label{sec:deployment}

We present case studies of using \sysname on two disparate networks, one a wide-area backbone and another a datacenter. In each case, using realistic test suites, we show that \sysname provides insight into what is and is not covered and how these insights help improve the test suites. 

\subsection{Case Study I: The Internet2 backbone}\label{sec:internet2}

\label{sec:internet2-overview}

Internet2 is a nation-wide network that connects over 60,000 US educational, research and government institutions. The routing design of Internet2 is typical of backbone networks. It has 10 BGP routers spread across the country. The routers are organized as a single autonomous system (AS), and they establish iBGP full mesh on top of internal reachability provided by the IS-IS protocol. The Internet2 routers connect to 279 external BGP peers, and heavily use route import and export policies. The import policy for an external peer has multiple policy statements, some specific to the peer and some shared within the same peer group. Peer-specific policies tend to specify a list of allowed prefixes from this peer, and others are used for sanity checking, preference setting, etc. Export policies are similarly structured. 

Internet2's configurations that we study have 96,672 lines (in Juniper's JunOS format) across all routers. Of these, \sysname's coverage computation considers 64,886 lines. The bulk of the unconsidered lines correspond to device management, IPv6, and IS-IS protocol.

We do not have the data plane state of Internet2, which is needed to run data plane tests. We approximate it using Route Views~\cite{routeviews}, a repository of BGP routes from over two hundreds ASes worldwide. 
This data helps approximate BGP messages that external peers of Internet2 send to it. Consider a peer with AS number $X$. If we find a prefix $P$ in RouteViews with AS-path $[A, X, Y]$, we assume that the peer sends $P$ to Internet2 with AS-path $[X, Y]$. The existence of AS-path $[A, X, Y]$ means that AS $A$ must have a route to $P$ with AS-path $[X, Y]$, which it announces to its neighbors. If we find multiple AS-paths for a prefix, we pick the one where $X$ is closest to the origin AS (the last entry in the path). 

We use these BGP messages that each peer sends to Internet2 as inputs to simulate Internet2's control plane using Batfish. The data plane state produced by this simulation is a coarse approximation of the real version, but it suffices to meet our goals of running data plane tests and characterizing configuration coverage. 

\subsubsection{Test suite coverage}\ \\
\label{sec:internet2-coverage}

To study how \sysname analyzes coverage for realistic test suites, we use the test suite proposed in Bagpipe~\cite{bagpipe}. It has three tests to validate Internet2's BGP configuration.

\newcommand{\BTE}{\textit{BlockToExternal}\xspace} 
\newcommand{\NM}{\textit{NoMartian}\xspace}
\newcommand{\RP}{\textit{RoutePreference}\xspace}
\newcommand{\SANITY}{\textit{SanityIn}\xspace}
\newcommand{\IFREACH}{\textit{InterfaceReachablility}\xspace}
\newcommand{\PSR}{\textit{PeerSpecificRoute}\xspace}
\begin{itemize}[itemsep=0pt]
    \item \BTE: ensure that BGP routes with BTE community are not announced to any external (eBGP) peer.
    \item \NM: ensure that incoming BGP messages from external peers for prefixes in the private address space ("Martian") are rejected.
    \item \RP: ensure that if multiple routes to the same prefix are accepted from multiple external neighbors, the selected route belongs to the most preferred neighbor. The neighbor's preference depends on commercial relationship~\cite{gao-rexford}.  \textit{Customers} are most preferred, followed by {\textit peers}, and then \textit{providers}.
\end{itemize}

We implemented these tests using Batfish. \BTE and \NM are control plane tests. \BTE evaluates all BGP export policies on a set of BGP routes carrying the BTE community and asserts that the result be rejection. We generate the test cases by sampling BGP routes from the data plane state and attaching the BTE community to them. \NM evaluates all BGP import policies on a set of BGP routes destined for Martian addresses and asserts that the results be rejection. \RP is a data plane test. It focuses on destination prefixes available via multiple neighbors and asserts that their local preferences reflect commercial relationship. We use CAIDA data~\cite{caida} to infer commercial relationship between Internet2 and its BGP neighbors.

After running this test suite on Internet2, we find that it covers only 26.1\% of configuration lines across all devices. Only a tiny fraction of configuration lines (0.5\%) are weakly covered, so we do not separate weak/strong  coverage for this case study; we will do that in the next one. 

To help understand what is and is not covered in more detail, \sysname enables network engineers to look at the data from multiple perspectives.
\Cref{fig:screenshot_aggregate} shows per-device coverage. We see notable variation across devices, from 11.8\% to 40.5\%. As we show below, the test suite has systematic gaps, and the cross-device variation stems from different devices having different fractions of covered configuration elements. 

\Cref{fig:coverage-original-tests} shows the coverage broken down by the type of configuration elements. For simplicity, we create four buckets of element types, as shown in the legend. The bottom bar shows the fraction of reachable configuration lines in each bucket. The "Test Suite" bar shows the covered fraction of those lines, and the top three bars show the coverage of individual tests. The total coverage of individual tests is $0.6$\%, $0.9$\% and $24.7$\% respectively. \BTE and \NM cover only one type of configuration element (routing policies), and even within this type, they cover a small fraction. \RP covered all four buckets but its overall coverage is still limited.

Finally, \sysname reports that 27.9\% of configuration lines are "dead code" that will never be exercised. They include defined BGP peer groups with no members and defined routing policies that are never used for any peer.\footnote{Per best practices, these lines should be deleted. Or, at a minimum, they should be tested lest someone start using an unused, erroneous policy. When it comes to testing, such lines can never be exercised by data plane tests, though control plane tests may be written for them. }

With 69\% of BGP configurations,  85\% of interfaces,  88\%  of routing policies, and  57\%  of route attribute match lists being completely untested, this test suite is clearly under-testing the network. This leaves the network vulnerable to bugs in untested configurations elements. Prior to \sysname, it was not possible for network engineers to get any insight into the quality of their test suite. It was also not possible for them to get help toward systematically improving tests. We demonstrate this test suite improvement process next.


\subsubsection{Coverage-guided test development}\label{sec:internet2-testdev}


\sysname's feedback enables a test suite development process that enables users to systematically improve coverage, which helps test more critical aspects of the network and prevent outages. This process is iterative. In each iteration the user first identifies specific testing gaps and then creates new tests to target those gaps. We demonstrate the process using three iterations that focus on different types of gaps.


\begin{figure}
    \centering
    \includegraphics[width=\columnwidth]{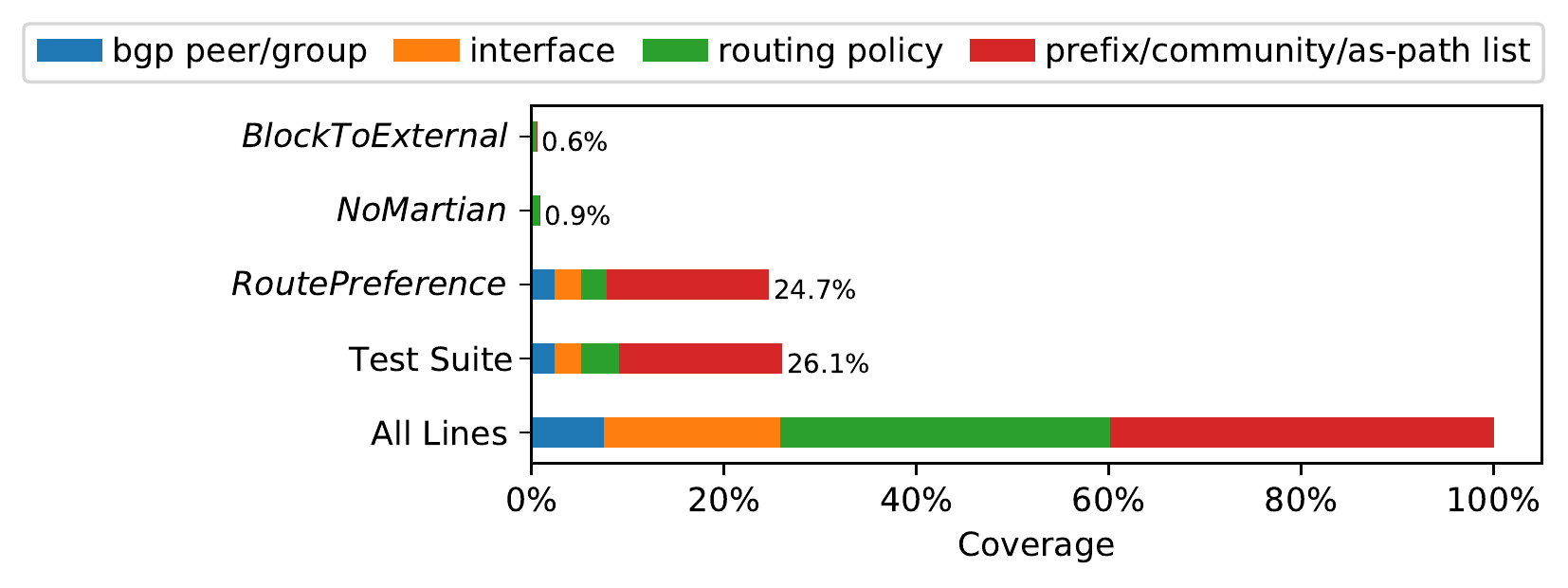}
    \vspace{-10pt}
    \caption{Coverage of the initial test suite broken down to each individual test and configuration type. }
    \label{fig:coverage-original-tests}
\end{figure}

\para{Iteration 1}
We saw that routing policy coverage of \NM test is low (\Cref{fig:coverage-original-tests}) despite that it checks the import policies for all external peers. 
To investigate, we look at the structure of Internet2 import policies and find that routers have a policy named \code{SANITY-IN} which is shared by the majority of external neighbors. \Cref{fig:screenshot_lcov} shows this policy with annotated coverage. Each router has an independent copy of this policy, but the copies and the coverage results are identical across routers. Of the five clauses in the policy, the clause \code{block-martians} starting at line 6,896 is the only clause that is covered. 
This coverage result confirms that the \NM test did its job, and more importantly, it revealed a systematic testing gap--the other four classes of forbidden routes are not being tested.

Once we know the gap, the solution suggests itself. We added a new test, \SANITY, to enforce that the other four classes of received BGP messages should be rejected. 
After adding this test, we used \sysname to confirm that this testing gap had been addressed. Routing policy coverage was improved by 0.6\% and all five terms of \code{SANITY-IN} were covered by the new test suite. 
The quantitative improvement is low because \code{SANITY-IN} is just one of many policies in the network. With feedback from \sysname, network engineers can identify testing gaps in other routing policies and add more tests in a similar way.\footnote{Automatic test generation based on coverage feedback will further help engineers. We will investigate this in the future.}


\para{Iteration 2}
BGP peer configuration coverage of \RP test in \Cref{fig:coverage-original-tests} is surprisingly low, given that all external BGP peers are supposed to be checked. Upon further investigation we find that the uncovered peers have permitted prefix-lists that do not overlap with other peers' lists, which left these peers untested.


We added a new test, \PSR, to check that BGP announcements received from external peers should be accepted if their prefixes is in a peer-specific prefix list. This test improved 
BGP peer coverage from 32\% to 46\%. The rest of untested BGP peers are either not allowed to send BGP routes to Internet2 or is intended for other internal use, such as monitoring and management. This test also improved prefix-list coverage from 45\% to 63\%. The remaining of untested prefix-lists are mostly (30\% out of 37\%) ones that are defined by never referenced.

\para{Iteration 3}
The low coverage of interface configuration in \Cref{fig:coverage-original-tests} reveals another testing gap. \RP is the only test in the initial test suite that checks interface configurations, and it only considers one category of interfaces--ones that are used to establish the tested BGP edges. 
Many other interfaces remain untested, including but not limited to ones that associate with untested BGP edges and other routing protocols, and the ones that are unused.

We added a new PingMesh-style~\cite{pingmesh} test, \IFREACH, to check that the IPv4 addresses assigned to interfaces should be reachable from each router in the network. This test increased interface coverage from 15\% to 53\%. The rest of untested interfaces do not have IPv4 addresses assigned.

\begin{figure}
    \centering
    \includegraphics[width=\columnwidth]{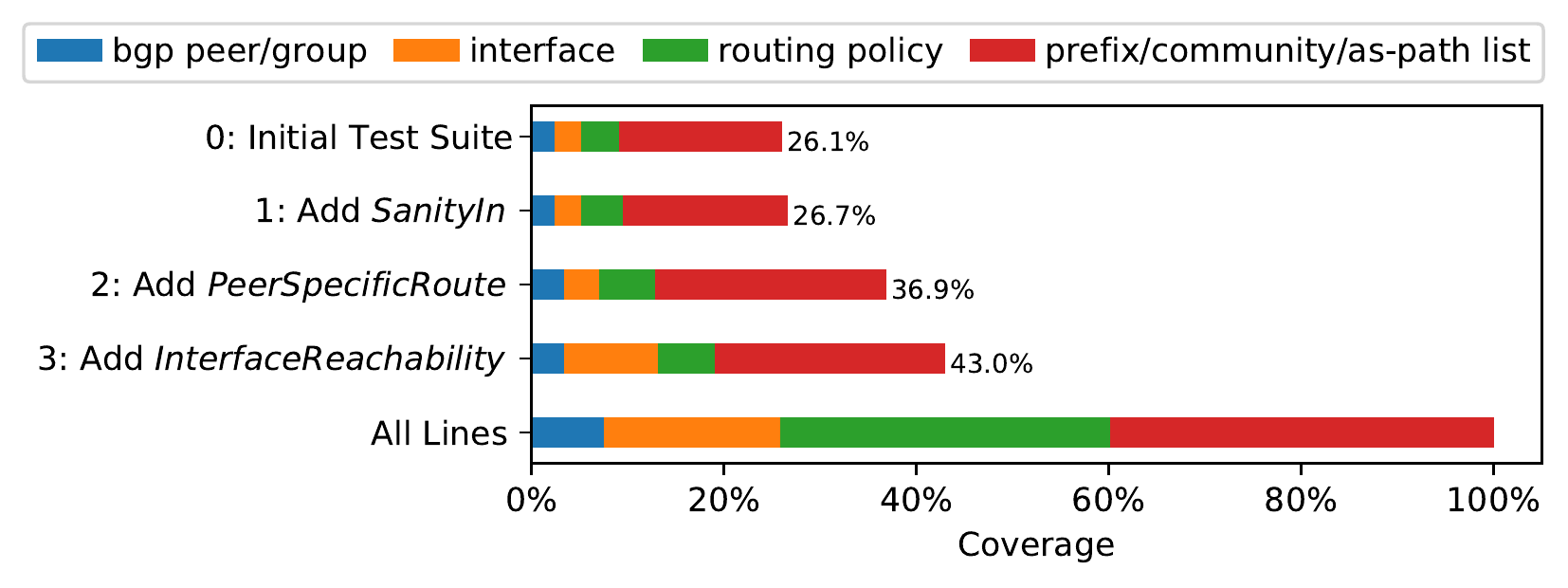}
    \vspace{-10pt}
    \caption{Coverage improvement with test suite iterations.}
    \label{fig:progression}
\end{figure}

\Cref{fig:progression} summarizes the coverage improvement for the three iterations of test improvement in our study. After only three iterations, the overall coverage was improved from 26\% to 43\%. 
This final coverage number is far from perfect, but our goal was not to develop the ideal test suite for Internet2; we wanted to  demonstrate how coverage information helps develop new tests. 
Networks are complex and we should not expect to get the job done with 6 tests. Many more tests are likely needed. With \sysname, network engineers now have a tool to develop new tests that meaningfully improve coverage.

\subsection{Case study II: Datacenter networks}\label{sec:dc}
We study the coverage for data center networks which have a different topology and routing design. We create synthetic fat-tree\cite{fattree} networks with routers across three tiers. The leaf routers at the bottom tier connect to hosts. Aggregation routers at the middle tier connect to leaf routers in a pod and to spine routers at the top tier. The spine routers connect to the wide area network (WAN). The WAN is not part of the tested network. 
Each leaf router is assigned a /24 prefix which is advertised inside the data center through eBGP. Spine routers receive a default route (prefix 0.0.0.0/0) from WAN via eBGP and propagate it to lower tiers. At each spine router, the entire address space of the network is summarized into a /8 prefix and is announced to WAN. Multipath routing (ECMP) is enabled with maximum number of paths set to 4. Routing policies are only configured at spine routers to white-list the default route received from WAN peers. We synthesize the configurations of these networks in Cisco IOS format.

We study a test suite of three tests inspired in prior works on data center network validation~\cite{rcdc,pingmesh}.

\newcommand{\DRC}{\textit{DefaultRouteCheck}\xspace} 
\newcommand{\PM}{\textit{ToRPingmesh}\xspace}
\newcommand{\EA}{\textit{ExportAggregate}\xspace}
\begin{itemize}[itemsep=0pt]
    \item \DRC: ensure that each router has the default route.
    \item \PM: ensure that each leaf router's assigned subnet is reachable from all other leaf routers.
    \item \EA: ensure that each spine router exports the aggregate route to WAN.
\end{itemize}
\begin{figure}
    \centering
    \includegraphics[width=\columnwidth]{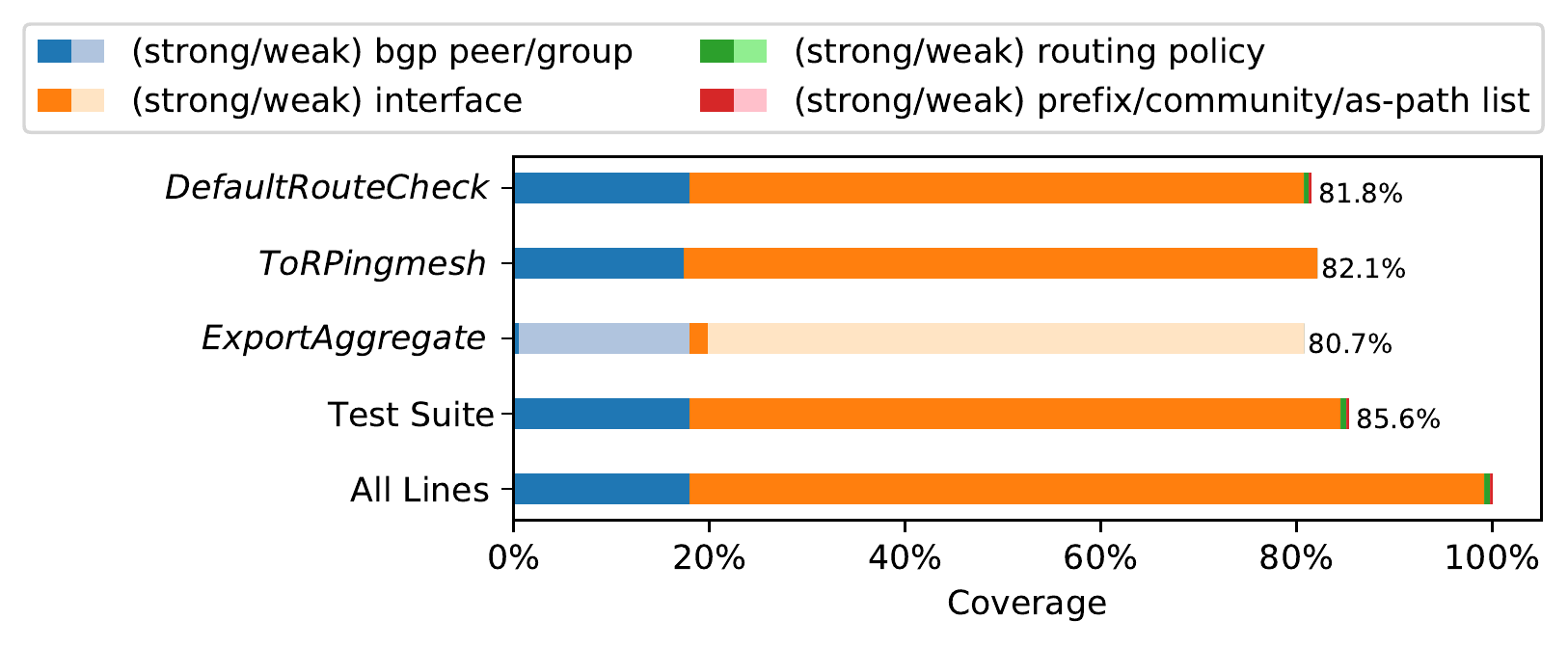}
    \vspace{-10pt}
    \caption{Coverage of synthetic datacenter network broken down to each individual test and configuration type.}
    \label{fig:fattree-coverage-original-tests}
\end{figure}
\Cref{fig:fattree-coverage-original-tests} shows the coverage result when the network has a total of 80 routers. Given the uniformity of the network and the test suite, coverage results are similar for other network sizes.
The total coverage of individual tests is $81.5$\%, $82.1$\% and $80.7$\%  respectively, and the three tests together cover $85.3$\% of configuration lines. We find that these tests cover largely the same configuration elements---interfaces and BGP peerings between the data center routers---despite checking for seemingly different network behaviors. This result indicates that seemingly distinct network tests can be redundant in terms of configuration testing. 

The coverage of \EA shows a large proportion of weak coverage. This is because a spine router has routes to all leaf routers, so that all leaf subnets contribute to the tested aggregate route, albeit weakly. Separating out weak coverage here avoids false negatives of testing gaps---the aggregate routes would be there even if some of the BGP peering or interfaces are misconfigured, therefore testing the aggregate routes provides a weaker endorsement for the covered BGP peerings and interfaces to be bug-free.

By looking at uncovered configuration lines reported by \sysname, we learn that most correspond to host-facing interfaces on leaf routers.  Adding tests that target those interfaces improves this test suite and eliminate testing gaps. We omit results of this iteration.



\begin{figure*}
    \centering
    \subfigure[Internet2.\label{fig:time_internet2}]{
    \includegraphics[height=.48\columnwidth]{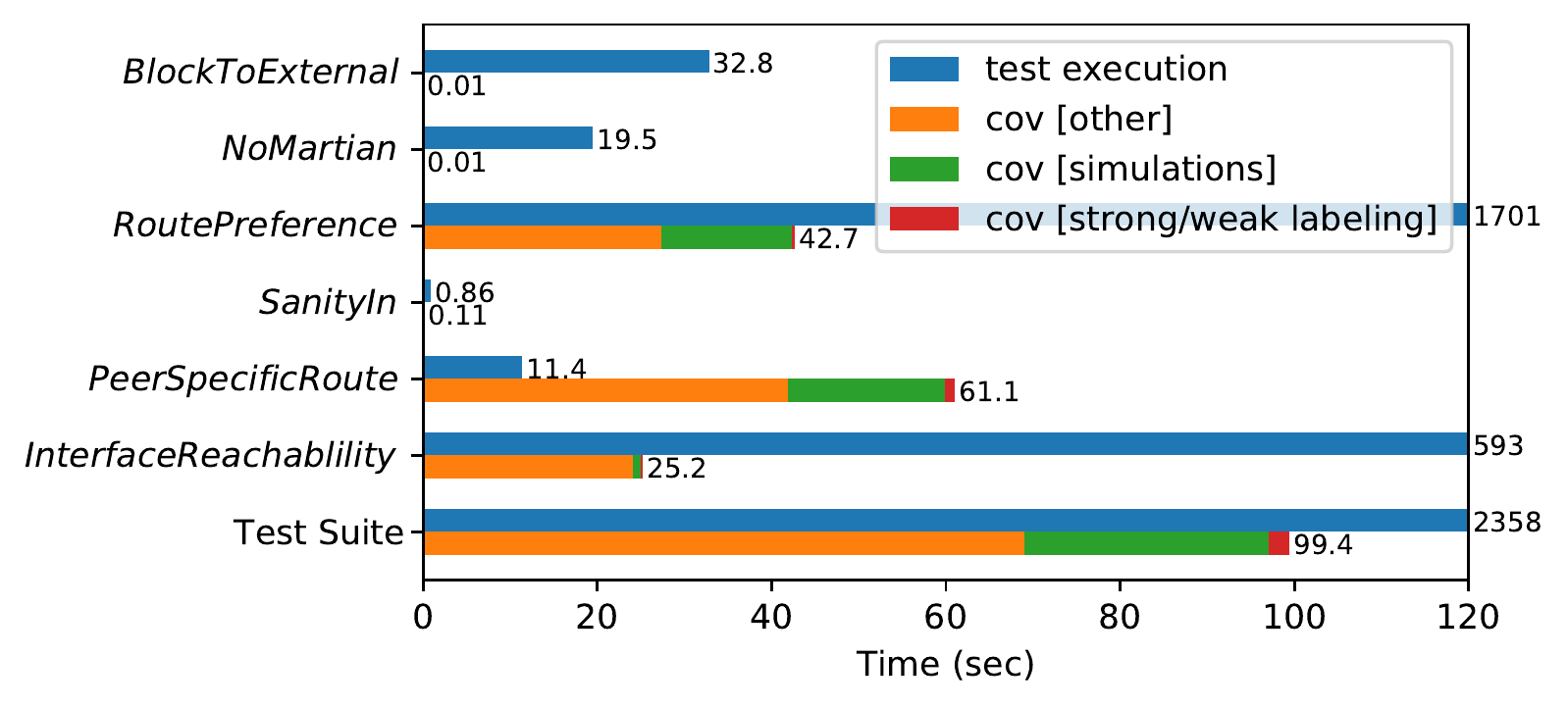}
    }
    \hfill
    \subfigure[Fat-tree networks.\label{fig:time_fattree}]{
    \includegraphics[height=.48\columnwidth]{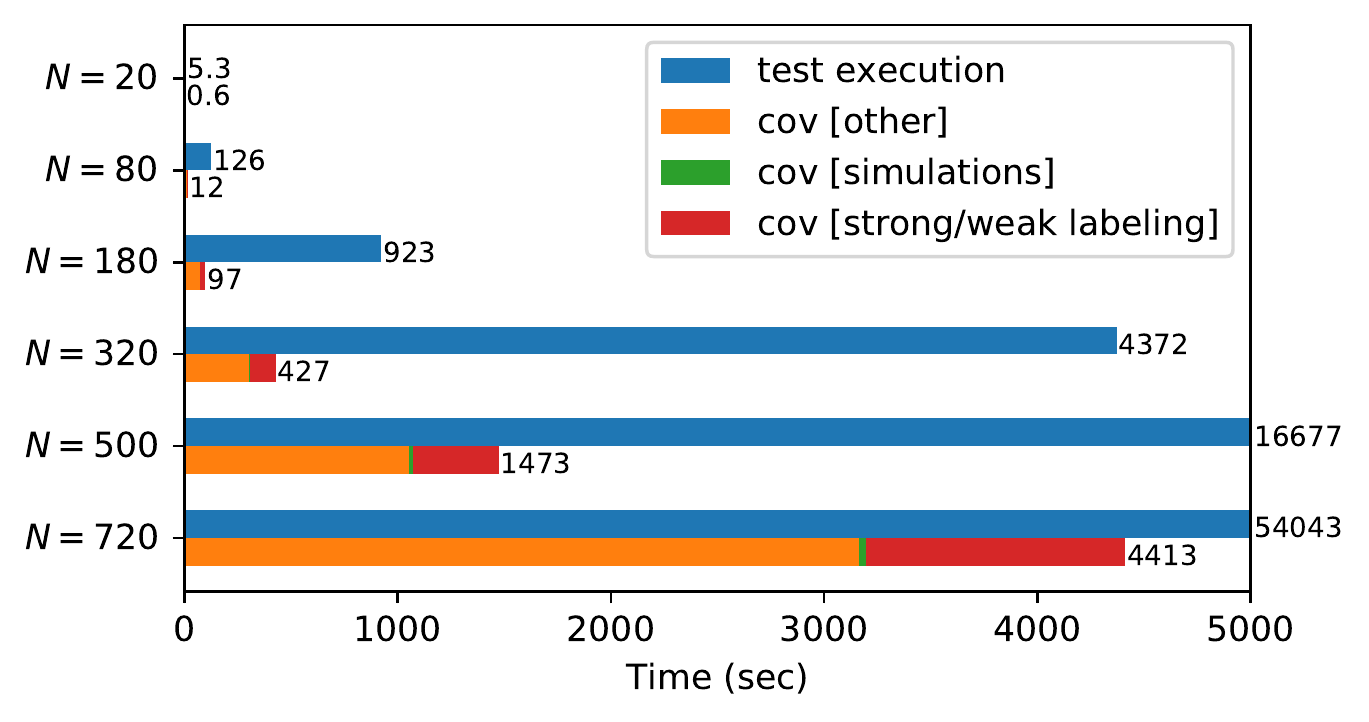}
    }
    \vspace{-10pt}
    \caption{Time to compute coverage.}
    \label{fig:performance}
\end{figure*}

\section{Performance Evaluation}\label{sec:performance}
We benchmark the performance of \sysname on both types of networks we studied above. 
Our test machine has two Intel Xeon CPUs (16 core each, 3.1 Ghz), 384 GiB of DRAM, and runs Ubuntu 18.04.

\Cref{fig:time_internet2} shows the time to compute coverage for each test in \cref{sec:internet2} and for the full test suite. It breaks out the time spent on simulations and strong/weak labeling, and, for reference, also shows the test execution time. We see that coverage computation is reasonably fast. The full test suite takes only 99.4 seconds. In comparison, the test execution takes 2,358 seconds. The total coverage computation time is less than the sum for individual tests because facts tested by multiple tests are tracked only once. The graph also shows that simulations and strong/weak labeling are a minority component, which means that most of the time is spent on walking the IFG and doing lookups in stable state for backward inference. 


\Cref{fig:time_fattree} shows test execution and coverage computation time for the test suite in \cref{sec:dc}, as a function of the data center network size. Coverage computation takes 4,413 $sec$ on the largest network, which has 2,040,624 RIB entries. This time is less than 9\% of the time to execute the test suite. While substantial, we deem it acceptable in practice. Configuration coverage analysis can be run in the background, as code coverage is often run. \sysname does not slow down test execution, which is on the critical path to finding configuration errors and updating the network.

However, time to compute coverage increases rapidly with network size. This is because the number of RIB entries grows quadratically and so does the number of vertices in the IFG. We find that the average time to materialize an IFG node does not change substantially because all computation is local to the node. The scaling trends suggest that to scale \sysname to much larger networks, we need a concurrent implementation of IFG materialization. Our current implementation is single-threaded (as Python interpreter is single-threaded).



\section{Comparison to Data Plane Coverage}\label{sec:dp_vs_cp}
\begin{figure*}[t!]
    \centering
    \subfigure[Internet2.\label{fig:cov_internet2}]{
    \includegraphics[height=.47\columnwidth]{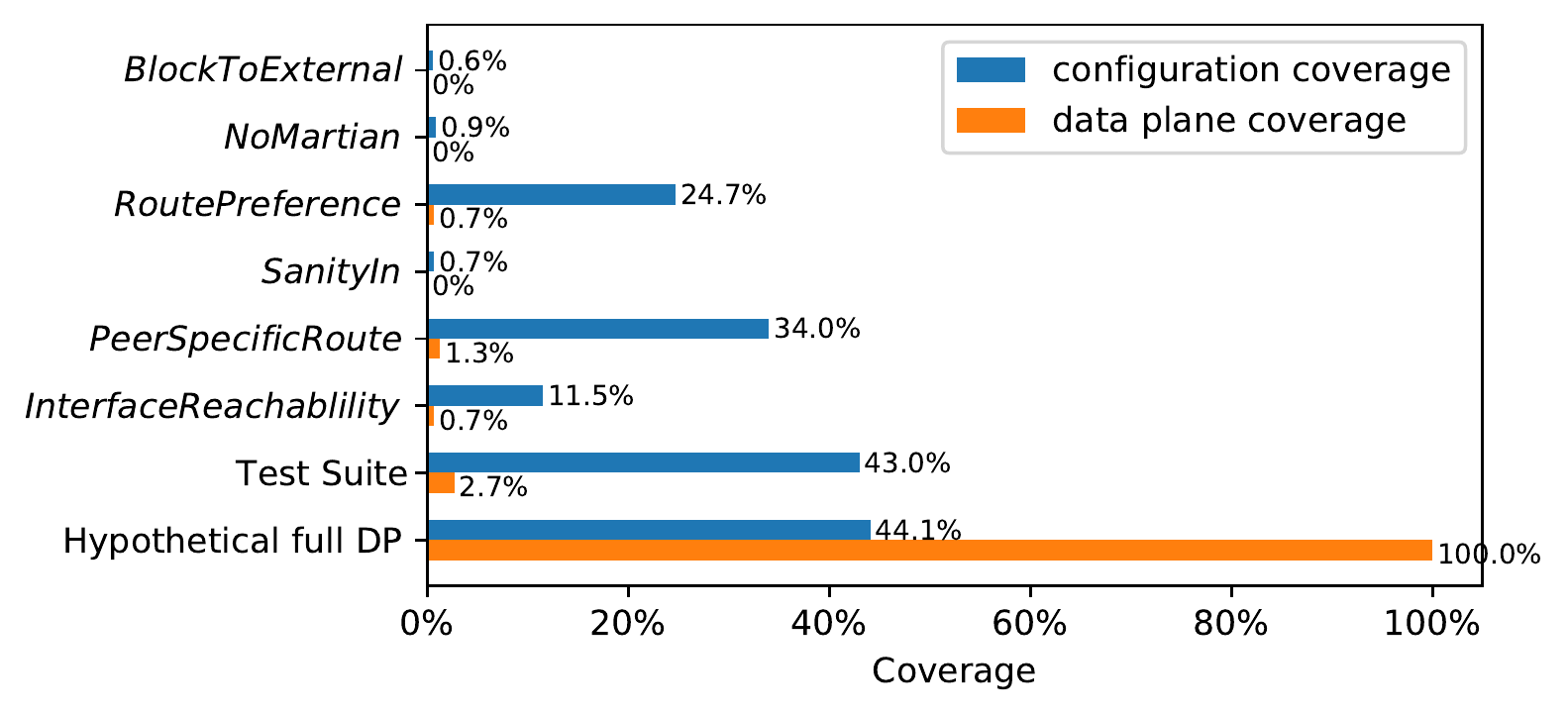}
    }
    \hfill
    \subfigure[Fat-tree with $k$=10.\label{fig:cov_fattree}]{
    \includegraphics[height=.47\columnwidth]{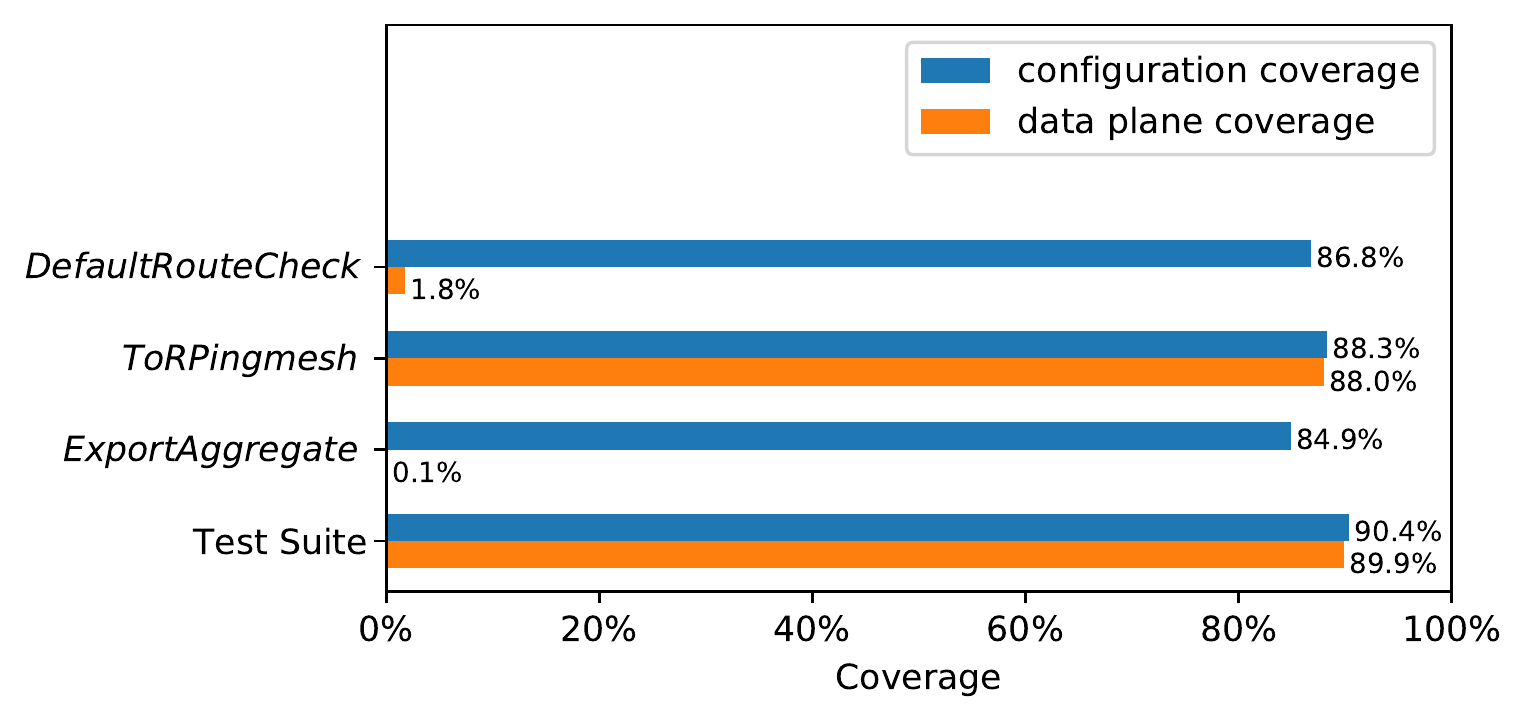}
    }
    \vspace{-10pt}
    \caption{Comparing control plane and data plane coverage.}
    \label{fig:dp_cp}
\end{figure*}

We demonstrate the unique value of control plane coverage by comparing it to data plane coverage. Following Yardstick~\cite{yardstick}, we quantify data plane coverage as the proportion of main RIB (forwarding) rules exercised.  \Cref{fig:dp_cp} shows the comparison for different cases. \Cref{fig:cov_internet2} shows the comparison for Internet2 for all tests in \cref{sec:internet2} and a hypothetical data plane test that inspects all main RIB rules. \Cref{fig:cov_fattree} shows the comparison for fat-tree tests in \cref{sec:dc}.

Besides the obvious advantage that only control plane coverage can support control plane tests---the graphs show 0\% data plane coverage for these tests---there are two main advantages to using control plane coverage to guide network test development.
First, it reveals testing gaps that can not be revealed by data plane coverage. Tests with high data plane coverage do not necessarily have high control plane coverage, as we can see in the last row of \Cref{fig:cov_internet2}. Covering 100\% of the data plane state covered only 41\% of the configuration. If the engineers were to improve the test quality under the guidance of only data plane coverage, they would not know that 59\% of the configurations remain untested. The reason of this disagreement is that some configuration lines are only exercised under specific environments (failures, routing messages). For instance, list-filtered route policies apply on BGP messages within a specific range, and will only be exercised when such messages appear in the environment. 

Second, testing more data plane state can sometimes be redundant in covering configurations, when the tests hit the same configuration elements. For example, the \DRC test in \Cref{fig:cov_fattree} has only 1.8\% data plane coverage because it only tests default routes, which is a small fraction of all main RIB routes. However, because correct propagation of default routes incorporates many BGP peerings and interfaces in the network, this test has extensive configuration coverage (87\%). The \PM test covers much more data plane state (88\%), but adding it atop \DRC has little value because this state is derived from almost the same set of configurations lines. 
We do not necessarily imply that engineers should drop one of these tests, as there may be other reasons to keep both. Our observations are about their value toward configuration coverage. 




\section{Related Work} 
\label{sec:related}

Our work builds on top of four lines of research.

\para{Code coverage}
We borrow from the software domain the idea of using code coverage to reveal testing gaps, quantify test suite quality, and help engineers improve their test suites~\cite{branch-coverage,mutation-coverage,overview-coverage}. Our coverage analysis techniques, however, are specialized to the operation of network configurations. 

\para{Data plane coverage}
Yardstick introduced data plane coverage metrics \cite{yardstick} that quantify the proportion of data plane elements such as forwarding rules and paths that are exercised by network tests. Configuration coverage goes further and maps tested data plane components to configuration elements that contribute to them. It provides more direct feedback because network engineers author configurations, not data plane state, and it supports testing of configuration elements that are not exercised by the current data plane state.  

\para{Network testing and verification}
A range of tools can analyse properties of network data and control planes~\cite{pingmesh,rcdc,hsa,batfish,hoyan,veriflow,deltanet,atomic,minesweeper,arc}.
\sysname borrows ideas from verification tools to concisely model the network, \EG, focusing on stable state and routing protocol instances~\cite{minesweeper,arc}. However, \sysname target a different problem---reveal what is tested vs enabling testing of new properties--and uses different techniques.

\para{Network provenance} 
Provenance systems can track causal dependencies of events in distributed systems.
Provenance systems like ExSPAN \cite{positiveprovenance} materialize provenance graphs by tracing system execution in forward direction.
Negative provenance systems can reason about missing events \cite{negativeprovenance} and materialize provenance graphs lazily using backward inference.
\sysname too uses a graph-based model. However, it is unique in terms of accommodating network configuration into a provenance model, and this model, tailored to the stable state assumption, is more succinct. Further, it combines backward and forward inference to overcome the limitations of using only one type of inference. 



\section{Summary}
\sysname is a tool that analyzes which configuration lines are tested by a suite of network tests. It uses an information flow model based on control plane semantics to track which configuration lines contribute to tested data plane state. It accounts for non-local and non-deterministic contributions, and for performance, it discovers the graph lazily. Our experiments showed that \sysname successfully reveals coverage gaps for real-world networks and test suites, and these tests can have surprisingly low coverage, e.g., 26\% of configuration lines for Internet2. They also showed how its feedback helps improve coverage.

This work does not raise any ethical issues.


\bibliographystyle{plain}
\bibliography{references}



\end{document}